\documentclass{article}

     \PassOptionsToPackage{square,numbers, compress}{natbib}

     \usepackage[final]{neurips_2020}

\usepackage[dvipsnames]{xcolor}
\usepackage[utf8]{inputenc} 
\usepackage[T1]{fontenc}    
\usepackage{hyperref}       
\usepackage{url}            
\usepackage{booktabs}       
\usepackage{amsfonts}       
\usepackage{nicefrac}       
\usepackage{microtype}      

\usepackage[pdftex]{graphicx}

\title{Randomized Overdrive Neural Networks}

\author{%
  Christian J. ~Steinmetz \\
  Queen Mary University of London\\
  \texttt{c.j.steinmetz@qmul.ac.uk} \\
  \And
  Joshua D.~Reiss \\
  Queen Mary University of London \\
  \texttt{joshua.reiss@qmul.ac.uk} \\
}

\begin{document}
 
\maketitle
\vspace{-0.3cm}

\begin{abstract}
By processing audio signals in the time-domain with randomly weighted temporal convolutional networks (TCNs),
we uncover a wide range of novel, yet controllable overdrive effects.
We discover that architectural aspects, such as the depth of the network, 
the kernel size, the number of channels, the activation function, as well as the weight initialization, 
all have a clear impact on the sonic character of the resultant effect, without the need for training. 
In practice, these effects range from conventional overdrive and distortion,
to more extreme effects, as the receptive field grows, similar to a fusion of distortion, equalization, delay, and reverb.  
To enable use by musicians and producers, we provide a real-time plugin implementation.
This allows users to dynamically design networks, listening to the results in real-time.
We provide a demo and code at \small{\url{https://csteinmetz1.github.io/ronn}}.
\end{abstract}

\section{Introduction}

Throughout the history of audio technology, engineers, circuit designers, 
and particularly guitarists, have searched for novel sonic effects as a result of clipping or distorting audio signals. 
These distortion effects were first discovered by pushing early guitar amplifiers beyond their operating range, 
or, in some cases, from the accidental damage to amplifiers or speakers \cite{shepherd2003distortion}. 
These pursuits are a clear example of creators taking advantage of the limitations of their tools for creative effect.
Distortion effects have permeated many genres such as blues, jazz, rock, and metal, and also play a central role in modern pop and hip-hop styles.
Whether it be vacuum tubes, diodes, integrated circuits, or software-based digital models,
it appears as if nearly all the methods of generating distortion-based effects have been exhausted.
We claim this may not be the case, as we will examine the potential of neural networks for audio signal processing to generate a new class of distortion-based effects. 

Neural networks are far from new. 
In fact, they arose in the same era that blues guitarists began their experiments with distortion \cite{schmidhuber2015deep}.
Yet, only more recently, following the emergence of modern deep learning approaches, 
have neural networks become feasible for audio signal processing \cite{purwins2019deep}.
Interestingly, these methods have shown to be successful in emulating the characteristics of amplifiers and distortion effects 
\cite{schmitz2018nonlinear, zhang2018lstm, damskagg2019distortion, martinez2019nonlinear}.
While these approaches have been successful in the emulation task, our aim deviates from these virtual analog effect modeling approaches.
In a similar spirit to the guitarists who used their amplifiers in a fashion unintended by the original designer,
we propose the apparent abuse of neural networks.
By utilizing randomly initialized networks as complex signal processing devices,
we aim to distort, transform, and warp audio signals for creative effect. 

\section{Method}
\subsection{Architecture}

We select a convolutional architecture as it provides a parameter efficient method for processing arbitrary length sequences.
From the perspective of audio signal processing, it can be considered a series of filters and nonlinear waveshapers.
The temporal convolutional network (TCN) formalizes the application of convolutional models operating on 1-dimensional sequences,
and outlines a set of design choices ideal for various sequence modeling tasks \cite{bai2018tcn}. 
Causal convolutions are a core component of this formulation, in which outputs are predicted considering only past values,
so information from the future does not "leak" into current predictions. 
Additionally, these networks are generally built to be fully convolutional, 
so they can process input signals of arbitrary length, and will produce an output signal with length proportional to the input length \cite{long2015fcn}.

\begin{figure}[htb!]
  \centering
  \includegraphics[width=0.4\textwidth]{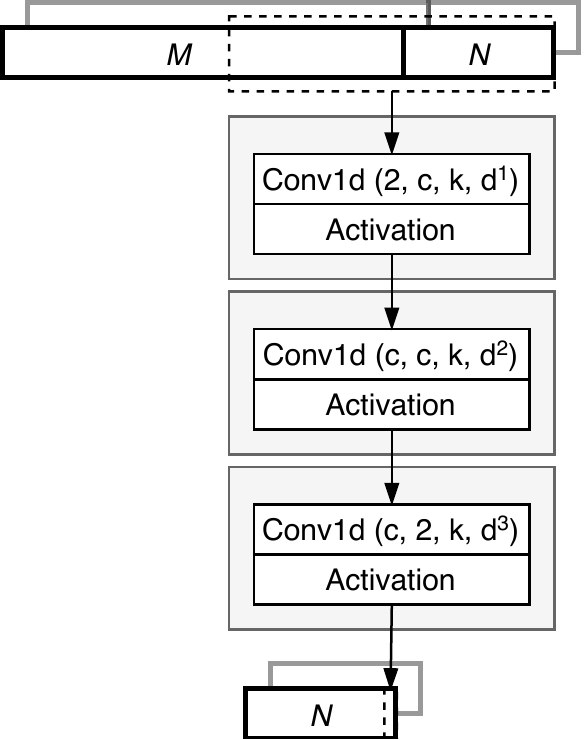}  
  \caption{Block diagram outlining the block-based processing as well as a 3-layer network in a stereo input-output configuration. 
  Each block within the network consists only of a 1-dimensional convolution followed by an activation function,
  where $c$ is the number of output channels, $k$ is the kernel size, and $d$ is the dilation growth factor.
  The look-back buffer is shown at the top, which consists of $N$ samples from the current input block, 
  concatenated with $M$ past input samples. The number of stored past samples $M$ is a function 
  of the receptive field of the TCN, shown in the dotted box, and the block size $N$. 
  $M$ is selected such that the entire buffer of size $M+N$ will produce an output of $N$ samples.
  Recall that since padding is not used, the output of each convolutional block will be smaller than the input. 
  When the receptive field is quite large, producing the output block (e.g. $\approx 10$ ms)
   may require processing multiple seconds of audio.}
  \label{fig:arch}
\end{figure}

With standard causal convolutions the receptive field grows linearly as the depth of the network increases. 
This makes it challenging to achieve models that are able to consider larger time contexts. 
To address this, the TCN incorporates dilated convolutions, which inserts zeros within the taps of the convolutional kernels, 
effectively increasing the size of the kernel without additional computation \cite{oord2016wavenet}. 
To increase the receptive field more rapidly, an exponentially increasing dilation factor is generally applied at each layer in the network. 
We omit residual connections since they are generally used at each layer to stabilize gradient flow, and we do not aim to train these networks \cite{he2016deep}.
This simplifies the overall architecture, which can then be viewed as a series connection of blocks containing
1-dimensional convolutions followed by a nonlinear activation, as shown in Figure \ref{fig:arch}. 

\subsection{Implementation}
\begin{figure}[h]
  \centering 
  \includegraphics[width=0.75\textwidth]{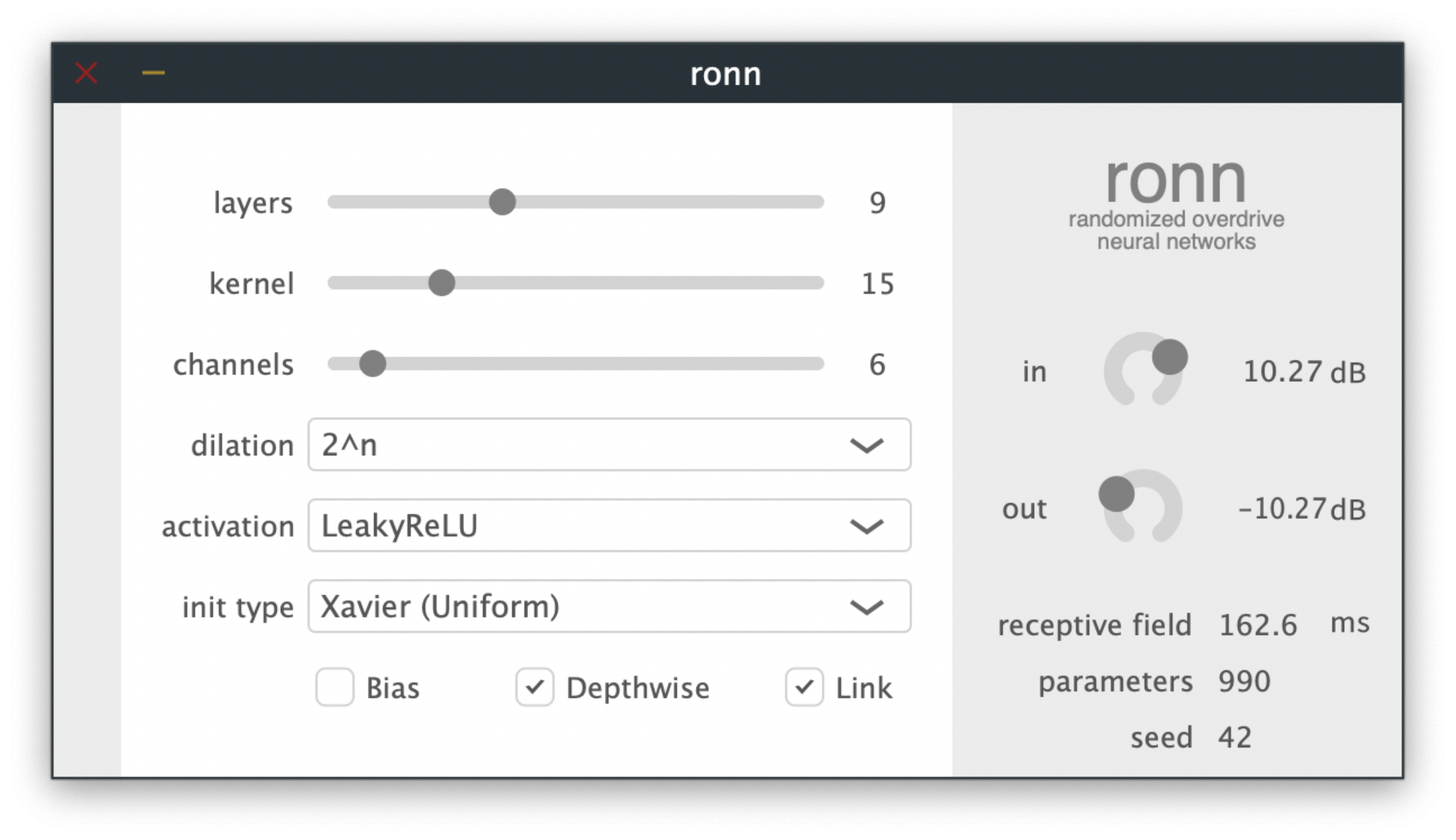}  
  \caption{The real-time plugin user interface featuring a series of sliders and selection boxes, 
  enabling users to dynamically construct various temporal convolutional network (TCN) architectures while listening to the results.
  When the user adjusts any of the on-screen controls, a callback will run, 
  constructing a neural network with the new architectural design. 
  On the bottom right, indicators show the receptive field in milliseconds, 
  the number of parameters in the network, and the global seed.}
  \label{fig:ui}
\end{figure}

For the real-time implementation, \texttt{ronn}, we utilize the JUCE framework\footnote{\url{https://github.com/juce-framework/JUCE}},
which enables us to create a VST/AU plugin for use in popular digital audio workstations (DAWs).
In order to construct the TCN models, we utilize PyTorch \cite{pytorch}, which features a C++ API. 
This enabled us to develop our own parameterized neural network module class that can be instantiated within the main JUCE plugin.
By connecting this class with the user interface, the on-screen controls, as shown in Figure \ref{fig:ui},
can be used to dynamically construct new networks, all in a paradigm that allows for real-time interaction. 

A challenge arises since these models are fully convolutional,
yet the plugin requires that all processing occur on a block-by-block basis. 
To address this, we construct a look-back buffer large enough so that the output sequence matches the length of the input block, 
as demonstrated in Figure \ref{fig:arch}. 
In practice, we found this approach produces no perceivable discontinuities at the frame boundaries. 
As expected, as the size of the receptive field increases, the computational load increases, causing the plugin to perform in less than real-time.
We introduce the ability to swap traditional convolutions with depthwise convolutions \cite{howard2017mobilenets},
which reduce the computational overhead by convolving $K$ filters with a single channel each. 
Using this approach, we were able to run models with larger receptive field, up to 4 seconds, in real-time on the CPU,
making a wider range of effects achievable on general purpose hardware utilized by musicians and producers. 

Using only a few layers, we achieve subtle to heavy distortion effects similar to traditional approaches, 
but by stacking more layers and expanding the kernel size, delay-like effects emerge. 
Using even larger receptive fields produces extreme temporal smearing, similar to reverberation.
Adjusting the shape of the nonlinearities also has a significant effect on the timbre of the distortion produced. 
We found that sigmoid activations often produced very harsh and gritty results, while ReLU activations \cite{nair2010rectified} produced fuzz-like effects.
Depending on the depth of the network, we found the weight initialization scheme also impacts the timbre of distortion. 
Since the TCN can produce any number of output channels, we can generate a stereo output signal, given only a mono input signal.
This also enables cross-channel interactions for stereo inputs. We found that these configurations often produced interesting spatialized results. 
Finally, a global seed control enables effect recallability and presets.

\section{Discussion}

We propose the use of randomly weighted TCNs as complex, time-domain audio signal processing devices. 
We find that adjusting various architectural aspects results in the ability to generate a wide range of compelling sonic effects.
This presents a new paradigm for designing audio effects. 
Instead of adjusting the controls of traditional processors, users can explore a wider space of effects by adjusting the architecture of a neural network.
While deep networks pose a challenge for real-time implementation due to significant compute overhead, 
we overcome this with some careful design choices. 
With a simplified interface, musicians and producers without machine learning experience can easily take advantage of these effects.
While we have only investigated feedforward architectures, 
it reasons that we could achieve a wider range of effects with the addition of recurrent pathways.

\section*{Broader Impact}
In this work, we applied existing neural network approaches for processing audio signals in a creative context.
Since we did not utilize any training data in this process, there is a relatively low risk of bias arising here. 
Potential biases in the processing of signals may arise from the randomized network weights,
which are sampled from various common distributions. 
Any current biases reflect the underlying characteristics of these generated effects. 
Future work could investigate potential biases in different weight initialization schemes used when processing different sources. 

\bibliographystyle{plain}  
\bibliography{references}
  
\newpage 
\section*{Supplementary materials}

\vspace{0.6cm}

\end{document}